\documentstyle{article} 
\include{epsf}
\begin{document} 

\title {{\bf Influence of a
low magnetic
field on the thermal diffusivity  of Bi-2212}} \vspace{5mm}
\author{S. Dorbolo(+)
and M. Ausloos (\dag)\\ \\ S.U.P.R.A.S\\ University of Li\`ege\\(+) Montefiore
Institute of electricity  B28\\ (\dag) Institute of Physics B5\\
B-4000 Li\`ege,
Euroland} \vspace{10mm} \maketitle \begin{abstract} The thermal
diffusivity of a
Bi-2212 polycrystalline sample has been measured under a 1T magnetic field
applied perpendicularly to the heat flux.  The magnetic contribution
to the heat
carrier mean free path has been extracted and is found to behave as a simple
power law.  This behavior can be attributed to a percolation process
of electrons
in the vortex lattice created by the magnetic field. \end{abstract}
\vspace{10mm}
\section{Introduction}

When a magnetic field is applied, it is well known that the superconductivity
properties collapse, e.g. (i) the critical temperature $T_c$
decreases \cite{TB},
(ii) the critical  current density decreases \cite{JB}, (iii) the electrical
resistance and (iv) the thermo-electrical power become finite below $T_c$
\cite{RB,SB}, (v) the thermal conductivity behaves anomalously
\cite{KB,Sousa}...
Indeed a magnetic field is known as a Cooper pair breaker. In fact the behavior
of the mean free path of electricity and heat carriers in e.g. high-T$_c$
superconductors materials is a widely discussed problem particularly
in presence
of a magnetic field.

Without a magnetic field, the thermal conductivity of these materials
exhibits a
hump below the critical temperature. The cause is not wholly clear nor really
understood: there are different view points, based on the prominence of an
increase of the mean free path of phonon, thus with reduced scattering
\cite{thconphon1,thconphon2,thconphon3,thconphon4} or  from an increase of the
mean free path of electrons \cite{illinois,NL,houssa,houssa2,houssa1,6,7}, or a
complicated combination of both due to the chemistry
\cite{thconphon5}. Note that the previously
quoted references are only a microsample of a huge literature on the
subject, but
not all reports (more than 300) can be quoted here.

On the other hand, in presence of any $finite$ applied magnetic
field, this mean
free path apparently decreases as indicated by the lowering of the  thermal
conductivity hump amplitude. The discussion of the findings pertain to the
understanding of the mixed state properties for d-wave (or s-wave as sometimes
thought) superconductors, and the role of phonons and electrons in a magnetic
field (see above references in appropriate cases, and also
\cite{3,8,9}). If they
are more or less well understood at high field and low temperature, their
interpretation is not so trivial in the regions where the various phase
transition lines merge into each other, thus near $T_c$ and at low field
\cite{2D3D}.

Since the thermal diffusivity is related to the thermal conductivity, the same
fundamental questions arise here, - less studies are found. The  thermal
diffusivity is known to be a good probe for measuring the heat
carrier mean free
path \cite{calzona,nous,ourpre} and even find out the relevance of Van Hove
saddle points  in High $T_c$ superconductors \cite{DoBou}.  Discussing the
thermal diffusivity behavior below and above $T_c$ in absence of a
magnetic field
has recently allowed us to distinguish the contribution of electrons
and phonons
in the dissipation process. On the other hand, comparison of the thermal
diffusivity behaviors with and without magnetic field below $T_c$ should allow
for extracting the contribution of the magnetic field.  This would give
interesting informations about electron-vortex
interactions.\footnote{ Calzona et
al. have measured the diffusivity in a magnetic field. However they used the
phonon dissipation model for explaining heat conduction features, and did not
extract  the magnetic contribution to the mean free path \cite{calzona}.  The
behavior of this contribution can give very interesting informations about the
vortex-heat carriers interaction below $T_c$. } Here we show that such a
contribution can be extracted, even close to $T_c$, and at rather low
field. The
experimental data are our own measurements on a polycrystalline Bi-2212 sample.
Despite its polycrystalline morphology, the advantage of such a
sample resides in
the possibility to obtain it as a long bar out of a pellet, whence allowing a
fine sensitivity of the measurements even at small field.  This compound and
sample are expected to be representative of most  high-T$_c$ superconductor
materials, in particular of their anisotropic (2D) nature.

In a simple kinetic model formalism, the thermal diffusivity $\alpha$
is defined
as the ratio between the thermal conductivity $\kappa$ and the volumic specific
heat $c$, and can be thus expressed through \begin{equation} \label{diff}
\alpha=\frac{1}{3} v \ell \end{equation} where $v$ is the velocity of the heat
carriers and $\ell$ the mean free path.  Assuming $v$ to be a constant in a
restricted range of temperature of interest, as here, the thermal
diffusivity is
thus an adequate (or direct) probe for measuring the mean free path of heat
carriers.  When a magnetic field is applied, $\ell$ may be rewritten as usual
within the linear superposition approximation as  \cite{ziman} \begin{equation}
\label{ell} \frac{1}{\ell}=\frac{1}{\ell_0}+\frac{1}{\ell_{mag}} \end{equation}
where $\ell_0$ and $\ell_{mag}$ are the mean free path of heat carriers
respectively without and with the influence of an applied magnetic field.  From
Eq.(\ref{diff}) and Eq.(\ref{ell}) \begin{equation}\label{alpha}
\frac{\alpha_0}{\alpha}=\frac{\ell_0}{\ell}=1+\frac{\ell_0}{\ell_{mag}}=
1+\frac{\alpha_0}{\alpha_{mag}} \end{equation} where $\alpha_0$ and
$\alpha$ are
the thermal diffusivities without and with magnetic field respectively and
$\alpha_{mag}$ is the magnetic contribution to the thermal diffusivity.

These formulae serve as a basis for the discussion of the vortex-heat carrier
interaction in the mixed phase near T$_c$. After a description of the synthesis
and of the characterization of the sample in Sect.2, the measurements of the
thermal diffusivity of the Bi-2212 sample are presented in Sect.3.
The magnetic
contribution to the mean free path is extracted and the results are then
discussed.

\section{Synthesis and experiments} Powders of Bi$_2$O$_3$, SrCO$_3$, CaCO$_3$,
CuO and PbO$_2$ were mixed mechanically together using an agathe
mortar starting
from a 2234 stoichiometry in order to obtain a 2212-BSCCO (Bi-2212)
stoichiometric composition. This mixture was decarbonated at 800$^o$C
for 20h and
melted in an alumina crucible at 1075$^o$C for 30 minutes in air. The
liquid was
quenched between two room temperature copper blocks to form a glass.
The pellets
were heated in oxygen atmosphere on a barium zirconate substrates at
860$^o$C for
50h \cite{so}.  Despite its polycrystalline morphology, the advantage of such a
sample is that it can be long.  This allows a better sensitivity for the
measurement and the effect of a small field can be as such observed.

The electrical resistivity curve is represented on Fig.1.  The superconducting
midpoint transition occurs around 85 K.  At high temperature, the
resistivity is
linear with a 10m$\Omega$.m resistivity at 0K and about 40m$\Omega$.m at 225 K.
When a magnetic field is applied, the resistivity transition midpoint
is shifted
towards lower temperature, ca. 80 K  for a 1.0 T field.

The thermal diffusivity has been measured as described in \cite{exp} and the
technique is  briefly recalled here.  A rod is first cut out from one of the
chemically characterized Bi-2212 pellet.  One of the extremities of the bar is
fixed to a heat sink, whilst the other is linked to a heather.  Three
thermocouples are set along the sample at equal distances from each
other. A heat
pulse is sent through the sample from one end and the change of temperature is
recorded by one of the thermocouples.  This operation is renewed 3
times, namely
once per thermocouple.  The signals recorded by the two extreme thermocouples
give the limit (''boundary'') conditions so as to compute the shape
of the signal
from the heat diffusion equation, at the middle thermocouple, for
different {\it
a priori} values of the thermal diffusivity.  The results of such calculations
are compared to the measured signal at the middle thermocouple.  The best fit
allows us to deduce the value of the thermal diffusivity at this temperature.

\section{Results and discussion}

In Fig.2, the thermal diffusivity is shown between 20 and 160 K.   The black
bullets ($\bullet$) represent the thermal diffusivity $\alpha_0$ without any
magnetic field.  As for the circles ($\circ$), they symbolize the
results with a
1.0 T magnetic field applied perpendicularly to the heat flux,
$\alpha$.  The two
curves are seen to be superposed on each other at high temperature.  Such a
magnetic field is indeed too small to create any visible effects on thermal
properties in the normal state.  On the other hand, the thermal diffusivity
differently behaves below the critical temperature with and without
the magnetic
field: in presence of a magnetic field the diffusivity is slightly lower.  Thus
the magnetic field shows its expected pair breaker role.  In the
framework of the
electronic model for heat transport, \cite{illinois,NL,houssa} that means that
the electron-electron scattering is enhanced, or in other words that electrons
have a decreasing mean free path.

The inset of the Fig.2 is the thermal diffusivity with and without the 1.0 T
magnetic field in a log-log plot.  This plot emphasizes that a break in the
slopes occurs at 85K, the critical temperature of the Bi-2212 phase.
This should
be expected from our previous report \cite{ourpre}.  The change in magnitude is
due to the sudden increase of the mean free path of electrons below
$T_c$.  They
are indeed less scattered by their counterparts since some electrons belong to
condensed Cooper pairs in this temperature range.  Thus the visible deviation
between the thermal diffusivity with and without a magnetic field  shows that
even a small magnetic field (1.0 T here) markedly acts on the thermal transport
in superconducting phase.

The magnetic contribution to the mean free path can be obtained from
Eq.(\ref{alpha}).   The results are shown in Fig.3, i.e. $\ell_{mag}/\ell_0$
versus the reduced temperature $\varepsilon=|T-T_c|/T_c$.  Notice that some
numerical smoothening data is necessary in order to reduce error bar
propagation.
It is seen that the normalized value $\ell_{mag}/\ell_0$ behaves as a power law
with an exponent found to be equal to $-0.5$.

This power law behavior and the exponent value itself remind us of
the Azlamazov-Larkin
law \cite{AL,varla} for the paraconductivity in its mean field
regime. The law is
usually studied between the onset temperature $T_o$ and $T_c$, thus
above $T_c$.
However the Azlamazov-Larkin law holds also below $T_c$ because of the scaling
hypothesis universality. The temperature region which is here above studied
extends much below the critical temperature and the precision of the data near
$T_c$ cannot be expected to lead to critical fluctuations studied {\it per se},
but the mean field exponent of superconductivity fluctuations might
be probed. It
can be shown that the same type of contribution exists in the electrical ($
\Delta \sigma$) and thermal ($\Delta \kappa$) paraconductivity
\cite{AuslVarla,Houssa3,Houssa4,Houssa5}. With the trivial change of
variables on
the temperature axis $T_c \rightarrow 0$ and $T_o$ $\rightarrow$ $T_c$
respectively, we can write $ \Delta \sigma \simeq \Delta \kappa \simeq \Delta
\alpha$, for the $parathermal$ $diffusivity$, the $\Delta$ notation  indicating
in both cases the deviation form the normal state behavior. Remembering that

\begin{equation} \label{assimil} \Delta \sigma_{3D} = \frac{e^2}{32
\hbar \xi(0)}
\varepsilon^{-1/2} \end{equation} for a $3D$ type of fluctuations
\cite{AL,varla}, and assuming that 1T is a low field such that  the
Azlamazov-Larkin law is still obeyed we can from the amplitude obtain
an estimate
of the shortening of the zero temperature coherence length $\xi$  between zero
and 1 T field. A simple numerical calculation leads to
${\l_0(0)}/{\l_{mag}(0)} =
0.938$.

As e.g. for neutron scattering processes in disordered 2D (magnetic) systems
\cite{cowley} the total inverse correlation length can be assumed to be the sum
of a strictly thermal term and a geometrical term linked (in that case) to the
disordered network, i.e. a relation similar to Eq.(\ref{ell}) for $\xi$.  This
analogy indicates that the found power law behavior can be interpreted as
resulting from a percolation process of heat carrier through a
complicated vortex
state \cite{3,SA}, and further justifies the analogy with the
Azlamazov-Larkin law
{\it for heat carriers in the mixed state} at low fields. One might also wonder
\footnote{ a referee comment to the first submitted version} why a
$3D$ behavior
(and exponent is found rather that the exponent, i.e. $\simeq -1.0 $
corresponding to a $2D$ behavior, since Bi-2212  is expected to have
an effective
dimensionality closer to 2 than 3. The  argument stems from the anisotropy
itself.   At the critical temperature itself, the coherence length diverges and
one expects a $3D$ behavior. When departing from $T_c$  the effective
dimensionality signature should appear, thus leading here to a $ 2D$ behavior.
However this lasts as long as the temperature is in the temperature
range limited
by the onset of the true critical regime, i.e. the Ginzburg-Levanyuk
temperature,
\cite{varla} and  by the Lawrence-Doniach temperature, in anisotropic systems
\cite{AuslVarla}. Away from the Lawrence-Doniach temperature, up to the onset
temperature, or at low temperature down to the temperature at which
the coherence
length saturates, one should recover a $3D$ regime \cite{2D3D}. In
Bi-2212, this
[Ginzburg-Levanyuk; Lawrence-Doniach] temperature range is estimated to be
50K, $\sim J_z=4.1$ meV \cite{houssa2}  the  exchange integral  along the
c-axis. The data where the
exponent $1/2$ is found is far away from the critical regime, and thus is the
signature of the geometrical disorder intrinsic to the polycrystalline system.

\section{Conclusion} The mixed state of high $T_c$ superconductors still has to
reveal many features and to be understood. Electrical and transport properties
contain the signature of the various vortex phases, and to distinguish them is
not so trivial.  Here the magnetic contribution to the mean free path of heat
carriers in a high-T$_c$ superconductor has been extracted in the low field and
in the near $T_c$ region. It has been found to increase below $T_c$ as a power
law characterized by a $1/2$ exponent.   This behavior is linked to a
percolation
process of the electron scattering in the vortex network
characterizing the mixed
state. We should emphasize that the above (temperature, field)
conditions are not
rather usual ones for probing the mixed state. If the diffusivity was
not so hard
to measure, its sensitivity to physical phenomena would be a bonus for
interesting conclusions. Such a remark may suggest new ways of probing whence
understanding the ($B,T$) phase diagram. \\ \vskip+5mm

{\bf Acknowledgments} \vskip+5mm Part of this work has been
financially supported
by the TMR network HPRN-CT-2000-0036 contract allowing some stay of SD at U.
Cambridge, UK during the writing of this report.  We would like also to thank
Prof. H.W. Vanderschueren for the use of MIEL equipments.  We want
also to thank
B. Robertz and R. Cloots for providing a large Bi-2212 good quality sample.

\newpage \vskip+10mm

\newpage 

\begin{figure}[htb]
\begin{center}
\leavevmode
\epsfysize=7cm
\epsffile{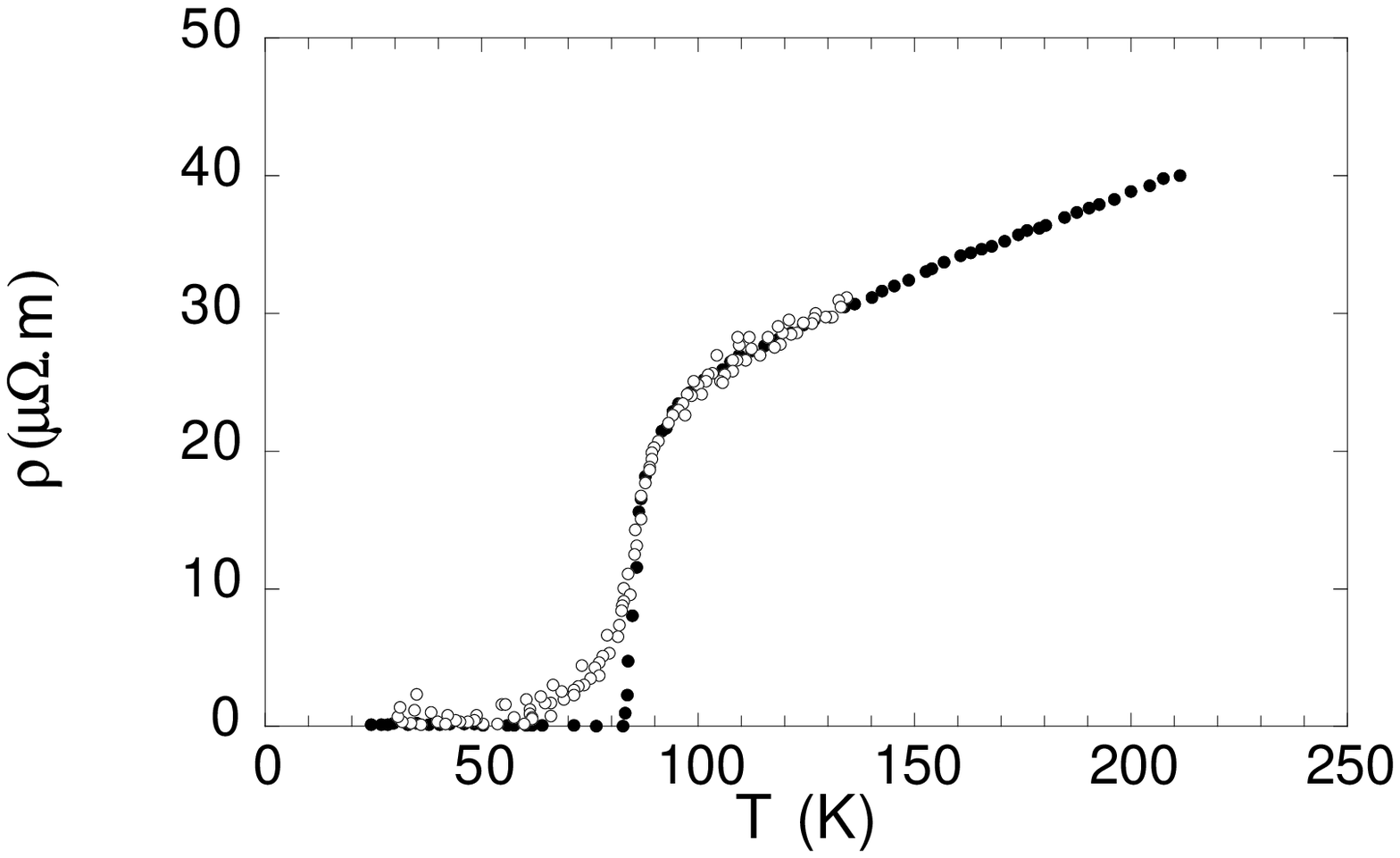}
\end{center}
\caption{Resistance of the
Bi-2212 sample with ($\bullet$) and without ($\circ$) an applied magnetic field
perpendicular to the current versus the temperature.}
\end{figure}

\begin{figure}[htb]
\begin{center}
\leavevmode
\epsfysize=6cm
\epsffile{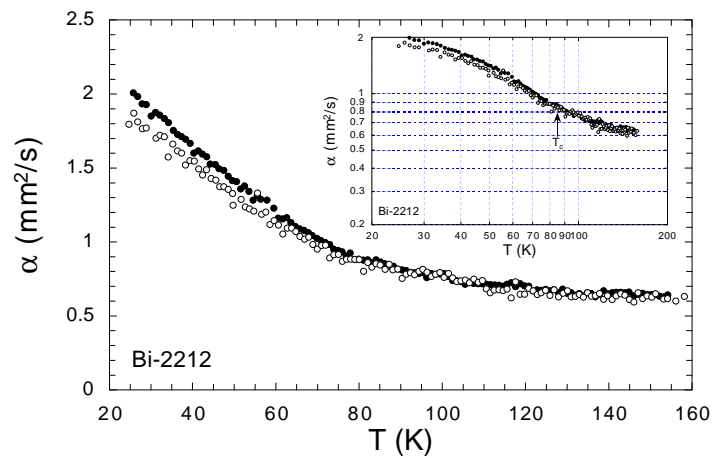}
\end{center}
\caption{Thermal diffusivity without and with a 1.0  T
applied magnetic
field versus the temperature, $\bullet$ and $\circ$ respectively.  The inset
represents the same quantities in a log-log plot.  The arrow indicates the
critical temperature. }
\end{figure}

\begin{figure}[htb]
\begin{center}
\leavevmode
\epsfysize=6cm
\epsffile{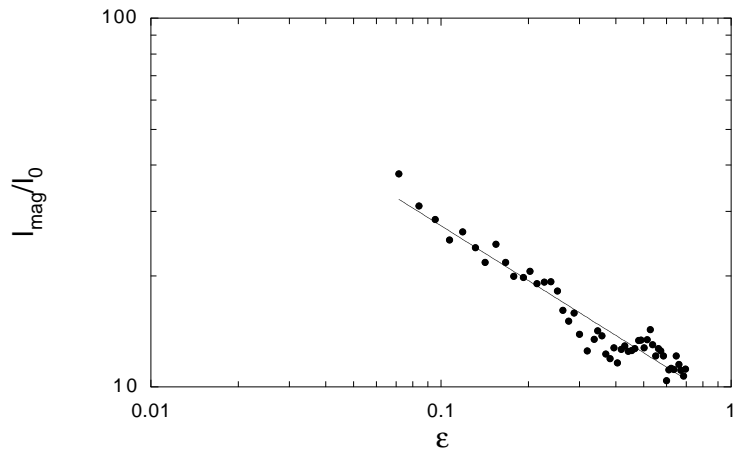}
\end{center}
\caption{ Normalized magnetic contribution of the mean free path plotted as a
function of the reduced temperature. }
\end{figure}

\end{document}